\documentclass[letterpaper,twocolumn,english,conference]{IEEEtran}
\usepackage[T1]{fontenc}
\usepackage{array}
\usepackage{amsthm}
\usepackage{graphicx}

\makeatletter

\pdfpageheight\paperheight
\pdfpagewidth\paperwidth

\providecommand{\tabularnewline}{\\}

\theoremstyle{plain}
\newtheorem{thm}{\protect\theoremname}
\theoremstyle{definition}
\newtheorem{defn}[thm]{\protect\definitionname}

\usepackage[lined,boxed,commentsnumbered,ruled]{algorithm2e}

\usepackage{babel}
\usepackage{subfigure}
\usepackage{array}
\usepackage{multirow}
\usepackage[usenames,dvipsnames,svgnames,table]{xcolor}
\usepackage{rotating}

\usepackage{balance}

\makeatother

\usepackage{babel}
\providecommand{\definitionname}{Definition}
\providecommand{\theoremname}{Theorem}

\begin{document}

\title{A Concise Network-Centric Survey of IP Traceback Schemes based on
Probabilistic Packet Marking}

\author{\IEEEauthorblockN{Matthias R. Brust}\IEEEauthorblockA{Center for Research in Cyber Security\\
Singapore University of Technology and Design, Singapore\\
matthias\_brust@sutd.edu.sg}\and \IEEEauthorblockN{Ankunda R. Kiremire}\IEEEauthorblockA{Center for Secure Cyberspace\\
Louisiana Tech University, USA\\
ark010@latech.edu}}
\maketitle
\begin{abstract}
Multiple \textit{probabilistic packet marking} (PPM) schemes for IP
traceback have been proposed to deal with \textit{Distributed Denial
of Service} (DDoS) attacks by reconstructing their attack graphs and
identifying the attack sources. 

In this paper, ten PPM-based IP traceback schemes are compared and
analyzed in terms of features such as convergence time, performance
evaluation, underlying topologies, incremental deployment, re-marking,
and upstream graph. Our analysis shows that the considered schemes
exhibit a significant discrepancy in performance as well as performance
assessment. We concisely demonstrate this by providing a table showing
that (a) different metrics are used for many schemes to measure their
performance and, (b) most schemes are evaluated on different classes
of underlying network topologies. 

Our results reveal that both the value and arrangement of the PPM-based
scheme convergence times vary depending on exactly the underlying
network topology. As a result, this paper shows that a side-by-side
comparison of the scheme performance a complicated and turns out to
be a crucial open problem in this research area.

\medskip{}
\end{abstract}

\begin{IEEEkeywords}
IP traceback, network security, distributed denial of service (DDoS),
probabilistic packet marking (PPM) 
\end{IEEEkeywords}

\section{Introduction}

Internet Protocol (IP) traceback is a method to deal with Distributed
Denial of Service (DDoS) attacks \cite{belenky,gao}. Using IP traceback,
sources of attack traffic can be identified from the network traffic
they generate. A prominent IP traceback technique for identification
of flooding style DDoS attacks is \textit{Probabilistic Packet Marking}
(PPM). In PPM-based IP traceback, network routers embed their own
identities in packets randomly selected from all the network traffic
that the routers process \cite{savage}. In the event of an attack,
the routers' identity markings present in the attack packets can be
used to reconstruct the \textit{attack graph} \textemdash{} the paths
taken by attack traffic \textemdash{} and establish its sources \cite{song,mesit2015secured}.
The technique of probabilistically marking packets for IP traceback
is the basis of many schemes hereafter referred to as \textit{PPM-based
schemes} \cite{song,yaar1,wong,paruchuri}. Multiple additional schemes
have been proposed with the purpose to increase the efficiency of
PPM-based schemes \cite{yaar1,paruchuri,goodrich,yan}. 

Tracing the paths of IP packets back to their origin, known as IP
traceback, is an important step in defending against DoS attacks employing
IP spoofing. IP traceback facilitates holding attackers accountable
and improving the efficacy of mitigation measures. The existing approaches
for IP traceback can be grouped into two orthogonal dimensions: packet
marking and packet logging. The main idea behind packet marking is
to record network path information in packets. In mark based IP traceback,
routers write their identification information (e.g., IP addresses)
into a header field of forwarded packets. The destination node then
retrieves the marking information from the received packets and determines
the network path. Due to the limited space of the marking field, routers
probabilistically decide to mark packets so that each marked packet
carries only partial path information. The network path can be constructed
by combining the marking information collected from a number of received
packets. This approach is also known as probabilistic packet marking
(PPM) \cite{savage}. PPM incurs little overhead at routers. However,
it requires a flow of marked packets to construct the network path
toward their origin.

In this paper, we present a concise network-centric analysis of a
selected set of PPM-based schemes. Ten PPM-based schemes are compared
in terms of features such as convergence time, the metrics, underlying
topologies, incremental deployment, re-marking, and upstream graph.
These schemes are PPM \cite{savage}, AMS \cite{song}, PPM-NPC \cite{tseng},
TMS \cite{ma}, FIT \cite{yaar1}, RPPM \cite{wong}, TPM \cite{paruchuri},
Randomize-and-link \cite{goodrich}, IDPPM \cite{yan}, and PBS \cite{ankunda}.
The schemes considered therein are by no means an exhaustive study
of all the PPM-based schemes in existence. However, the collection
of schemes is large enough to show the discrepancy in both the metrics
and underlying topologies as well as the inadequacy of the topologies
that make their direct comparison difficult \cite{kiremire2014topology,kiremire2014using}.

Thus, this paper shows that the direct performance comparison of the
schemes is complicated or not feasible at all. As our analysis show
the reasons for this is (a) many schemes utilize different metrics
to measure their performance and, (b) the many schemes are simulated
on different kinds of underlying network topologies, the majority
of which do not provide an adequate abstraction of the topology of
the Internet or focus on different characteristics of it \cite{wong,tseng,ma}. 

In detail, the underlying topologies are typically tree-structured
with a single path from an attacker to the victim. In contrast, the
topology of the Internet exhibits alternative routes that make it
both resilient and scalable. Both the disparity in metrics and underlying
topologies, as well as the inadequacy of the topologies, raise questions
about the performances of these schemes in the Internet and the comparability
between them. This shows that there is a need to evaluate and compare
the schemes on common and appropriate networks. The results of this
evaluation can then be used to determine which schemes are the most
promising candidates for Internet deployment.

In summary, the contribution of this paper is threefold: (a) an analysis
of PPM-based schemes is presented, (b) providing a taxonomy for PPM-based
IP traceback schemes, and (c) an analytical model to argue that different
underlying topologies inhibit a direct comparison between the performance
of the PPM-based schemes.

\begin{figure}[!tbp]
\centering{}\includegraphics[width=1\columnwidth]{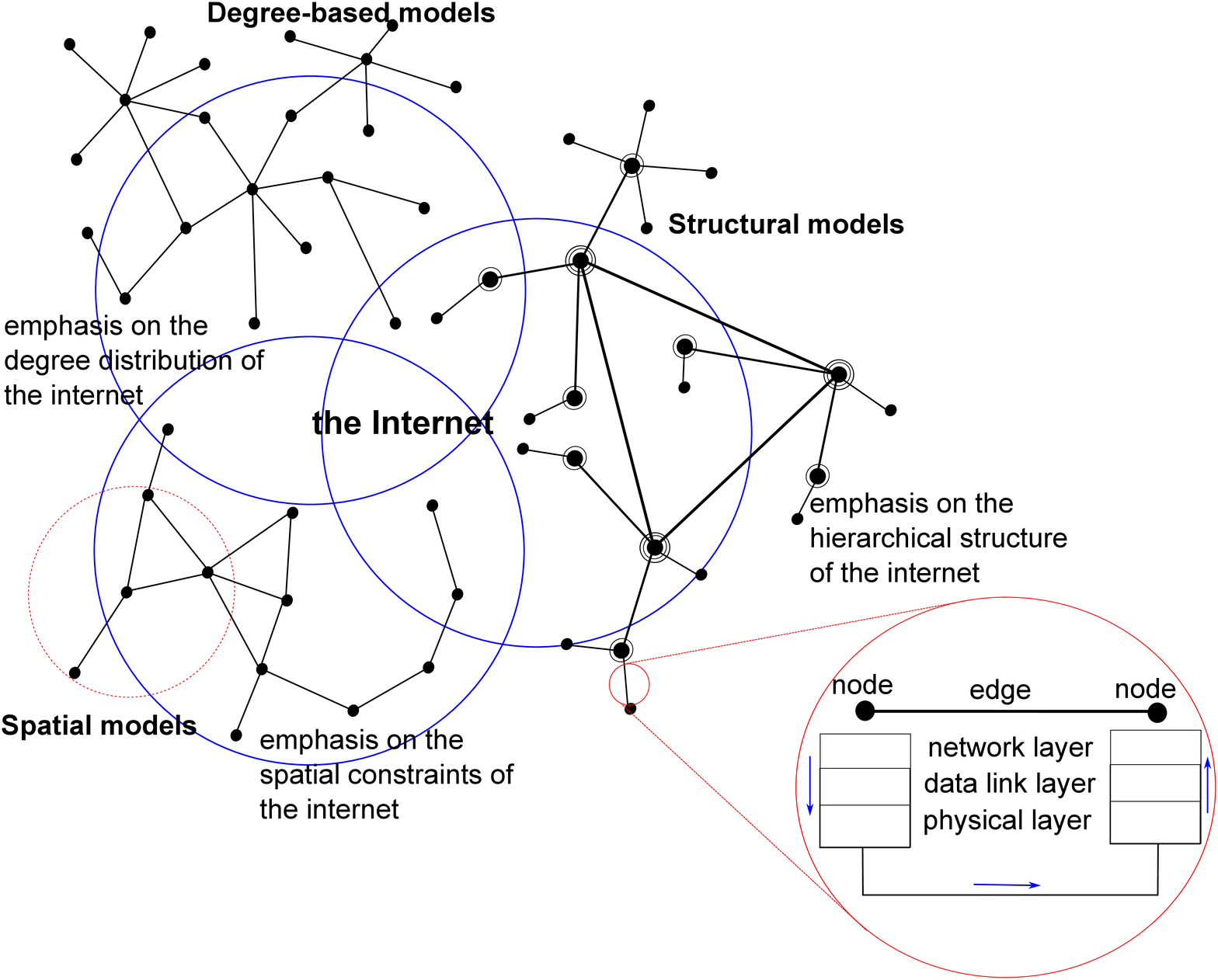}
\caption{Internet topology can be captured by a variety of models which include
spatial, structural and degree-based models. Each model emphasizes
different properties of the Internet and can be used to evaluate network
protocols when employed in the Internet. The nodes in these topologies
represent devices operating at the Internet layer of the TCP/IP model
or the network layer of the OSI model (e.g. routers, switches, hosts).
Two nodes are connected by an edge if Internet traffic can be directly
transmitted between them without being forwarded by any intermediate
nodes.}
\label{Internet} 
\end{figure}
The remainder of this paper is arranged as follows. Related work,
in which the selected set of PPM-based protocols is discussed, is
presented in Section \ref{sec:Related-work}. The analysis of PPM-based
schemes as well as additional background is presented in Section \ref{sec:Protocol-Analysis},
including considerations on definitions, metrics, and underlying topologies.
The theory and system model behind our work is analyzed in Section
\ref{analysis}. We discuss outcomes of this paper and conclude in
Section \ref{sec:Conclusion}.

\section{Related work\label{sec:Related-work}}

In this section, we describe and discuss a set of PPM-based IP traceback
schemes for which we will provide an analysis in Section \ref{sec:Protocol-Analysis}. 

PPM-based schemes consist of a \textit{marking scheme} and a \textit{reconstruction
procedure}, and are based on the assumption that large amounts of
traffic are used in a (D)DoS attack \cite{savage}. In their original
work, Savage et al. \cite{savage} propose that the PPM marking scheme
is employed at all times in all the routers in the network, while
the reconstruction procedure is employed by the victim in the event
of an attack. The marking scheme ensures that every router embeds
its own identity in packets randomly selected from the packets the
routers process during routing. Since a large number of packets is
received in an attack, there is a considerable chance that a victim
will have received packets with markings from all the routers that
were traversed by the attack packets. The victim then employs the
reconstruction procedure which uses the received marked attack packets
to map out the \textit{attack graph} \textemdash{} the paths from
the victim to the attackers. The total number of received packets
required to trace the attackers is referred to as the scheme's \textit{convergence
time}.

Multiple PPM-based schemes have since been introduced. One example
is the \textit{Tabu Marking Scheme} (TMS) \cite{ma}. The author points
out that PPM is prone to information loss as a result of \textit{re-marking}.
Re-marking occurs when a router randomly selects a packet which already
has marking information from an upstream router, and consequently
overwrites this information. TMS tackles this problem by ensuring
that their marking scheme forfeits the marking opportunity in the
event that the randomly selected packet contains previous marking
information. As a result, they report lower convergence times than
PPM for DDoS attacks.

Another example is the \textit{Prediction Based Scheme} (PBS) which
also avoids re-marking \cite{ankunda}. However, in contrast to TMS,
the PBS marking scheme ensures that the router information is embedded
in the next available packet if the randomly selected packet already
has marking information. The PBS marking scheme requires extra space
cost of 1 bit compared to PPM. Additionally, the reconstruction procedure
utilizes both legitimate and attack traffic to reconstruct the attack
graph.

Wong et al. \cite{wong} present \textit{Rectified Probabilistic Packet
Marking} (RPPM). They point out that the reconstruction procedure
used in PPM \cite{savage} and \textit{Advanced and Authenticated
Marking schemes} (AMS) \cite{song} has an imprecise termination condition.
They present a precise termination condition that enables complete
attack graph reconstruction within a user-specified level of confidence.

Multiple additional schemes have been proposed with the purpose to
increase the efficiency of PPM \cite{yaar1,paruchuri,goodrich,yan}.
Some of these schemes are analyzed in Table \ref{taxonomy}. The table
compares ten PPM-based schemes in terms of features such as convergence
time, the metrics, underlying topologies, incremental deployment,
re-marking, and upstream graph. The schemes considered therein are
by no means an exhaustive study of all the PPM-based schemes, but
the selection of schemes is large enough to show the discrepancy in
both the metrics and underlying topologies as well as the inadequacy
of the topologies that make their direct comparison difficult.

The feature \textit{incremental deployment} refers to whether the
scheme would be successful if the marking scheme is deployed on a
fraction of the routers in the network. Only a few schemes explicitly
state that they would be successful when partially deployed \cite{savage,song,yaar1,goodrich}.

\textit{Re-marking} refers to whether the marking scheme at a router
permits the overwriting of previous edge or router information in
a packet. The majority of the considered schemes permit re-marking
of packets \cite{savage,song,yaar1,wong,paruchuri,goodrich,yan}.

\textit{Upstream graph} refers to whether a scheme requires a previously
obtained map of the network to successfully trace the specific path
taken by attack traffic. Some of the works address how such a map
can be obtained to aid in attack graph reconstruction \cite{song,yaar1,ankunda}.

Other features such as \textit{metrics}, \textit{convergence time},
and \textit{underlying topology} are discussed in subsequent sections.

It is important to point out that PPM-based schemes are not the only
proposed approaches to IP traceback \cite{belenky,gao}. Alternatives
include packet logging \cite{sung}, specialized routing \cite{stone},
Internet control message protocol (ICMP) traceback \cite{bellovin},
deterministic packet marking \cite{belenky1} and hybrid approaches
which combine different traceback techniques \cite{yangtifs}, or
combine traceback with anomaly detection \cite{xiangtifs}.

\section{Protocol Analysis\label{sec:Protocol-Analysis}}

In this section, an analysis of the discussed PPM-based schemes from
Section \ref{sec:Related-work} is provided. This section is categorized
according to definitions and expressions, 

\subsection{Definitions and expressions}

The performance of traceback schemes is commonly measured by the \textit{convergence
time}, also known as the \textit{reconstruction time}. The convergence
time is defined as the total number of received packets required by
the reconstruction procedure to return the complete attack graph \cite{savage}.
The convergence time is dependent on the marked packet distribution
$P$, the number of attackers $n$, and attack graph $G$ as below.
\begin{defn}[Convergence time]
The \textit{convergence time} $C$ is defined as $C=f(P_{i},n,G)$
where $G(v,e)$ is a graph made up of a set $v$ of nodes and a set
$e$ of edges on which $M$ is implemented.
\end{defn}

\begin{defn}[Attack path length]
The \textit{attack path length} $d$ is the number of edges (hops)
that the attack traffic traverses between an attacker $A$ and the
victim $V$.
\end{defn}

\begin{defn}[Marked packet distribution]
The \textit{marked packet distribution} $P_{i}$ is described as
the marked packets received by the victim from the $i^{th}$ attacker
$A_{i}$ given $i\in[1,n]$ where $n$ is the number of attackers.
We define $P_{i}=f(p,d,M)$ as it depends on the marking probability
$p$, the marking scheme $M$, and attack path length $d$.
\end{defn}
An alternative measure of the performance of a traceback scheme is
the correctness of the attack graph $G_{ret}$ that the reconstruction
procedure returns when compared to the actual attack graph $G_{act}$. 

This is especially important in marking schemes where the router IP
address is broken into fragments because of space constraints in the
packet header as in \cite{savage,song,yaar1,paruchuri}. In these
works, a fragment of the router's IP address, instead of the entire
address, is transmitted in the packet header, which requires an extra
function of combining the fragments during the reconstruction procedure.
This introduces the possibility of incorrect recombination where fragments
from routers in $G_{act}$ are combined to identify an innocent router
in $G_{act}'$, whereby $G'$ refers to the complement of $G$, as
being part of $G_{ret}$. Ideally, the returned attack graph should
be identical to the actual attack graph, i.e. $G_{ret}=G_{act}$.
\begin{defn}[False positives]
\textit{False positives} are nodes incorrectly identified as belonging
to the attack graph. Given an attack graph $G_{act}$ and a reconstructed
attack graph $G_{ret}$, the false positives are identified as all
members of the set $\{G_{ret}\cap G_{act}'\}$.
\end{defn}

\begin{defn}[False negatives]
\textit{False negatives} are nodes in the attack graph that are not
identified as such by the reconstruction procedure. Given an attack
graph $G_{act}$ and a reconstructed attack graph $G_{ret}$, the
false negatives are identified as all members of the set $\{G_{act}\cap G_{ret}'\}$.
\end{defn}
The expressions for convergence time shown in Table \ref{taxonomy}
are the analytical bounds for $C=f(d,p)$ given $G$ is the \textit{Single
Path, Single attacker} (SP/SA) topology and the reconstruction procedure
has no prior knowledge of the underlying topology. The SP/SA topology
occurs when all the attack traffic originates from a single source
and takes the same path to the victim. The convergence time expressions
shown in Table \ref{taxonomy} are derived using the coupon collector's
problem \cite{feller}. In \textit{randomize-and-link} \cite{goodrich},
$l$ is the number of fragments that each router splits its address
into, and $H_{nl}$ is the $nl^{th}$ harmonic number. The schemes
in \cite{song,yaar1,paruchuri,yan} do not provide mathematical modeling
of their convergence times in the SP/SA topology.

Savage et al. \cite{savage} and Tseng et al. \cite{tseng} point
out that the ideal marking probability $p$ depends on the attack
path length $d$. In PPM \cite{savage}, the marking probability is
$p=0.04$. This value of $p$ is based on the report that the path
length $d\leq25$ for most Internet paths \cite{carter}. This value
of $p$ is maintained in AMS \cite{song}, TMS \cite{ma}, RPPM \cite{wong},
and PBS \cite{ankunda}.

Wong et al. \cite{wong} present a reconstruction procedure in which
they count the number of packets required to construct a fully connected
attack graph, and then count extra packets to guarantee with a certain
level of confidence that all edges in the attack graph are in fact
accounted for. To ensure uniformity across the schemes considered
in our simulations, we use a simplified version of their reconstruction
procedure in which the number of packets required to account for all
edges in the complete attack path is counted.

A large variety of graphs are used as the underlying topologies in
the considered protocol performance analysis. The majority of uses
tree-structured graphs $G_{tree}=\{v,e;|v|=|e|+1\}$ \cite{savage,wong,tseng,ma,goodrich,ankunda}. 

\begin{table*}[!t]
\newcolumntype{X}{>{\centering\arraybackslash}m{1.8cm}}
\caption{A comprehensive comparison of PPM-based IP traceback schemes showing
the variety of metrics employed, and the topologies used for the evaluation
of those schemes. The table provides a comparison of 10 different
PPM-based schemes over their features. These features include \textit{convergence
time}, \textit{metrics} used for evaluation of the schemes, whether
they require prior knowledge of the \textit{upstream graph} to correctly
identify attackers, whether they can be \textit{incrementally deployed},
and \textit{underlying topology}. The convergence time expressions
presented in the table are for a DoS scenario assuming no prior knowledge
of the network topology while the underlying topology shows the network
topologies used in the evaluation of the schemes. These topologies
include \textit{single path, single attacker} (SP/SA), \textit{single
path, multiple attacker} (SP/MA), and \textit{multiple paths, multiple
attacker} (MP/MA). The metrics shown in this table are described in
Table \ref{metrics}}
\label{taxonomy} \centering \scalebox{0.90}{ %
\begin{tabular}{|m{2.5cm}|c|c|c|X|c|X|m{4.5cm}|}
\hline 
\multicolumn{1}{|c|}{\textbf{Scheme}} & \textbf{Year} & \textbf{Convergence time} & \textbf{Metrics} & \textbf{Incremental deployment} & \textbf{Re-marking} & \textbf{Upstream graph} & \multicolumn{1}{c|}{\textbf{Underlying topology}}\tabularnewline
\hline 
\hline 
\multirow{2}{*}{\textbf{PPM} \cite{savage}} & \multirow{2}{*}{2001} & \multirow{2}{*}{$\leq\frac{\ln(d)}{p(1-p)^{d-1}}$} & \multirow{2}{*}{1} & \multirow{2}{*}{yes} & \multirow{2}{*}{yes} & \multirow{2}{*}{no} & \multirow{2}{*}{SP/SA (max. 30 hops)}\tabularnewline
 &  &  &  &  &  &  & \tabularnewline
\hline 
\textbf{AMS} \cite{song} & 2001 & \textit{undetermined} & 1,4,5 & yes & yes & yes & Traceroute data set (103402 destinations, 2000 attackers)\tabularnewline
\hline 
\multirow{2}{*}{\textbf{PPM-NPC} \cite{tseng}} & \multirow{2}{*}{2004} & \multirow{2}{*}{$\leq\frac{\ln(d)+0.58}{p}$} & \multirow{2}{*}{7} & \multirow{2}{*}{no} & \multirow{2}{*}{no} & \multirow{2}{*}{no} & \multirow{2}{*}{SP/SA (10 hops)}\tabularnewline
 &  &  &  &  &  &  & \tabularnewline
\hline 
\multirow{2}{*}{\textbf{TMS} \cite{ma}} & \multirow{2}{*}{2005} & \multirow{2}{*}{$\leq\frac{\ln(d)}{p(1-p)^{d-1}}$} & \multirow{2}{*}{1,2} & \multirow{2}{*}{no} & \multirow{2}{*}{no} & \multirow{2}{*}{yes} & \multirow{2}{*}{Binary tree (6 hops, 32 sources)}\tabularnewline
 &  &  &  &  &  &  & \tabularnewline
\hline 
\textbf{FIT} \cite{yaar1} & 2005 & \textit{undetermined} & 1,3,6 & yes & yes & yes & Skitter map (174409 hosts, 5000 attackers)\tabularnewline
\hline 
\textbf{RPPM} \cite{wong} & 2008 & $<\frac{\ln(d)}{p(1-p)^{d-1}}$  & 5,10,11,12  & no  & yes  & no  & SP/SA, binary tree, random tree network (15, 100, 500, 1000 nodes)\tabularnewline
\hline 
\multirow{2}{*}{\textbf{TPM} \cite{paruchuri}} & \multirow{2}{*}{2008} & \multirow{2}{*}{\textit{undetermined}} & \multirow{2}{*}{1,4,8,9} & \multirow{2}{*}{no} & \multirow{2}{*}{yes} & \multirow{2}{*}{yes} & \multirow{2}{*}{Skitter data (avg. 18 hops)}\tabularnewline
 &  &  &  &  &  &  & \tabularnewline
\hline 
\textbf{Randomize-and-link} \cite{goodrich} & 2008 & $<\frac{nlH_{nl}}{p(1-p)^{d-1}}$  & 7,11,12  & yes  & yes  & no  & Binary tree (10 hops)\tabularnewline
\hline 
\multirow{2}{*}{\textbf{IDPPM} \cite{yan}} & \multirow{2}{*}{2010} & \multirow{2}{*}{\textit{undetermined}} & \multirow{2}{*}{12,13} & \multirow{2}{*}{no} & \multirow{2}{*}{yes} & \multirow{2}{*}{no} & \multirow{2}{*}{SP/SA (20-32 hops)}\tabularnewline
 &  &  &  &  &  &  & \tabularnewline
\hline 
\textbf{PBS} \cite{ankunda} & 2012 & $\leq\frac{\ln(d)}{p}$  & 1,2,7  & no  & no  & yes/no  & SP/SA, SP/MA, MP/MA, 50 node network, 100 node network\tabularnewline
\hline 
\end{tabular}} 
\end{table*}

\subsection{Metrics}

In this section, we describe the metrics used in the considered PPM-based
schemes (as shown in Tables \ref{taxonomy} and \ref{metrics}) in
terms of the definitions given in the preceding section.

\subsubsection{Convergence time metrics}

The convergence time $C$ is the fundamental measure of the speed
and success of any PPM-based scheme and it is dependent on many factors.
A variety of metrics has been used to measure this dependence with
respect to these various factors.

\textit{Convergence time versus attack path length} shows how $C$
varies with $d$. This is normally done for a simple topology $G_{tree}$,
with $n=1$ but is done for larger values of $n$ in \cite{song,ma,wong,paruchuri,ankunda}.

\textit{Convergence time versus number of attackers} is a measure
of how $C$ varies with $n$ for a specified $G$. This metric is
considered in \cite{wong,goodrich,ankunda,tseng}.

\textit{Convergence time versus marking probability} is used by Wong
et al. \cite{wong} to investigate the relationship between $C$ and
marking probability $p$ for a topology $G_{tree}$, where $n=1$,
$d=3$ and show that large $p$ values actually increase the convergence
time while very low $p$ values result in large values of $P_{i}'$.
This relationship is investigated by Goodrich \cite{goodrich} for
$G_{tree}$, $n=1000$, $d=10$. He finds that the ideal marking probability
is $p_{ideal}\approx\frac{1}{d}$.

\textit{Convergence time versus number of routers} is a description
of the relationship between convergence time $C$ and the size $|v|$
of a given graph $G$. This metric is used in \cite{goodrich} for
$|v|\in\{50,100,250,500,1000\}$, in \cite{wong} for $|v|\in\{15,50,100,500,1000\}$,
and in \cite{yan} for $|v|\in[3,30]$.

\textit{Reconstruction time in seconds versus the number of attackers}
is used in \cite{song} to measure the variation of $C$ with $n$.
However, $C$ is translated from packets to time in seconds. As a
result, this metric is dependent on the system on which the victim
is running the reconstruction procedure, as well as the bandwidth
and latency of the topology on which the simulation is being carried
out. In \cite{wong}, the authors investigate the relationship between
reconstruction time in seconds and the number of nodes $|v|$ in the
attack graph.

\subsubsection{False positive and false negative metrics}

Some researchers have investigated the various factors that affect
the accuracy of the traceback process.

\textit{False positives versus false negatives} is a measure of how
$|\{G_{ret}\cap G_{act}'\}|$ varies with $|\{G_{act}\cap G_{ret}'\}|$.
This form of a \textit{Receiver Operating Curve} (ROC) is used to
evaluate the scheme in \cite{yaar1}.

\textit{False positive versus number of attackers} investigates the
variation of the number of false positives $|\{G_{ret}\cap G_{act}'\}|$,
with $n$ for $G=G_{tree}$. This is considered in \cite{song,paruchuri}.

\textit{False negative versus number of packets} shows how the false
negatives $|\{G_{act}\cap G_{ret}'\}|$ vary with the number of received
packets without any false positives. This plot is used in \cite{yaar1}
with $n=100$ to show that the accuracy improves with an increase
in received packets.

\begin{table}[!t]
\global\long\def\arraystretch{1.5}
 \caption{This table shows metrics used in the considered PPM-based schemes
and their indices from Table \ref{taxonomy}.}
\label{metrics} \centering \scalebox{0.95}{ %
\begin{tabular}{|c|l|}
\hline 
\textbf{Index}  & \textbf{Metric definition}\tabularnewline
\hline 
\hline 
1  & Convergence time vs. attack path length\tabularnewline
\hline 
2  & Number of marked packets vs. distance from victim\tabularnewline
\hline 
3  & False positives vs. false negatives\tabularnewline
\hline 
4  & False positives vs. number of attackers\tabularnewline
\hline 
5  & Reconstruction time in seconds vs. number of attackers\tabularnewline
\hline 
6  & False negatives vs. number of packets\tabularnewline
\hline 
7  & Convergence time vs. number of attackers\tabularnewline
\hline 
8  & Fraction of distinct markings vs. number of attackers\tabularnewline
\hline 
9  & Probability a packet is unmarked vs. path length\tabularnewline
\hline 
10  & Successful rate vs. traceback confidence level\tabularnewline
\hline 
11  & Convergence time vs. marking probability\tabularnewline
\hline 
12  & Convergence time vs. number of routers\tabularnewline
\hline 
13  & Overhead on routers vs. number of nodes\tabularnewline
\hline 
\end{tabular}} 
\end{table}

\subsubsection{Additional metrics}

There are additional metrics that do not fall into the convergence
time, or false positive and false negative categories. These metrics
are generally used to understand the extra dynamics involved in the
traceback process.

\textit{Fraction of distinct markings versus number of attackers}
measures the proportion of $P_{i}$ with distinct markings as a function
of $n$ at a given value of $d$. This metric is used in \cite{paruchuri}.

\textit{Number of marked packets versus distance from victim} is a
description of the distribution of $P_{i}$. It shows how $P_{i}$
varies with router distance $l$, where $l\in[0,d]$. This was evaluated
for $G_{tree}$ with $n\in\{1,32\}$ in \cite{ma}, and with $n\in\{1,3,5\}$
in \cite{ankunda}.

\textit{Probability that a packet is unmarked versus path length}
is used in \cite{paruchuri} to show the variation of $P_{i}'$ with
$d$ for $M\in\{PPM,TPM\}$.

\textit{Successful rate versus traceback confidence level} is used
by Wong et al. \cite{wong}. They present a reconstruction procedure
that uses a traceback confidence level to determine when to halt the
traceback process. They present this metric, which is a measure of
how the proportion of times $\{G_{ret}=G_{act}|G_{act}=G_{tree},n=1,d=3\}$
varies with the traceback confidence level for $p\in\{0.1,0.5,0.9\}$.

\textit{Overhead on routers versus number of nodes} is used in \cite{yan}
to show the overhead cost on the routers as a function of the number
of nodes $|v|$ in the attack graph $G$. The overhead on the routers
is measured in terms of the number of markings performed.

In summary, Tables \ref{taxonomy} and \ref{metrics} expose the large
number of metrics that are used to measure the performance of the
mentioned PPM-based schemes. Coupled with the differing settings used
to test these metrics, this makes direct comparison of different schemes
problematic.

\subsection{Categories of PPM schemes}

Despite their large number, PPM-based schemes have similar underlying
algorithms in their marking schemes. The underlying algorithm is responsible
for how the packets in which the router identities are embedded are
selected. For example, the majority of the considered schemes exhibit
underlying algorithms in which all routers randomly select packets
with equal probability $p$ \cite{savage,song,yaar1,wong,paruchuri,goodrich,yan}.
The schemes in this category are prone to re-marking. We refer to
this category as the re-marking category of PPM-based schemes. In
the other category of schemes, the routers' packet selection process
is only partially random. The underlying algorithms in this category
prohibit the overwriting of previous router information and as a result
exhibit performances that are notably different from the re-marking
category \cite{ma,ankunda,tseng}.

We select three representative marking schemes: PPM \cite{savage}
to represent the re-marking category, and TMS \cite{ma} and PBS \cite{ankunda}
to represent the non-re-marking category. The analytical models for
these three schemes are markedly different from each other, even for
equal marking probability, because of the differences in the schemes'
underlying marking algorithms. The performance of any PPM-based scheme
can therefore be compared to either one of these schemes, or a combination
of them.

Because of re-marking in PPM, the victim typically receives more markings
from close-by routers than from distant routers. The chance of receiving
a marked packet from a router $l$ hops away is given by the geometric
distribution expression $p(1-p)^{l-1}$. This is because a received
marked packet indicates that that packet was selected by a router
(with probability $p$), and not selected (with probability $1-p$)
by all ${l-1}$ subsequent routers. The analysis for PPM can therefore
be applied to any scheme where the markings from distant routers are
rarer than markings from close-by routers.

In TMS, the decision to forfeit a marking opportunity if the packet
is previously marked means that markings from routers distant from
the victim are more prevalent than markings from closer routers. The
chance of receiving a marked packet from a router $l$ hops away is
given by $p(1-p)^{d-l}$ where $d$ is the attack path length. This
is because a received marked packet indicates that that packet was
selected by a router (with probability $p$), after not being selected
(with probability $1-p$) by all ${d-l}$ previous routers. This analysis
can be applied to all schemes in which markings from distant routers
are more prevalent than markings from close-by routers.

In contrast to TMS, the PBS marking scheme compensates for the missed
marking opportunities. Therefore the chance of receiving a marking
from a router $l$ hops from the victim is given by $p$ for any router
in the path. This analysis can be applied to all schemes in which
the markings from the routers are equally prevalent regardless of
their distance from the victim.

These three schemes therefore provide an adequate basis to understand
the impact of the network topology on other PPM-based schemes.

\subsection{Underlying topologies\label{Underlying topology section} }

The performance of a PPM-based traceback scheme should be evaluated
on the Internet itself. However, because of the very large-scale dynamic
and heterogeneous structure of the Internet attempts to carry out
empirical protocol evaluation are expensive and inflexible \cite{calvert}.
As a result, researchers resort to simulations implemented on underlying
topologies which are considered to be simplified abstractions of the
topology of the Internet \cite{calvert,tang,faloutsos,brite2}. An
underlying topology is represented by a graph $G(v,e)$ consisting
of nodes $v$ and edges $e$ where the nodes represent either devices
with routing capability or end hosts. An edge between any two nodes
means that traffic can be directly transmitted between those two devices
(cf. Fig. \ref{Internet}) \cite{calvert}.

As shown in Table \ref{taxonomy}, a variety of underlying topologies
have been used to evaluate the performance of PPM-based schemes. During
set up, the marking algorithm is implemented in the nodes (routers)
in the graph. To simulate the attack, packets are transmitted from
one or more nodes (representing the \textit{attackers}) to one specific
node (representing the \textit{victim}). A reconstruction procedure
is then implemented at the victim to map out the attack graph $G_{act}$.
The attack graph consists of all nodes and edges in the underlying
topology that were directly involved in transmitting the attack packets.
The underlying topologies used in PPM-based schemes range from simplistic
to complex.

The \textit{single path, single attacker} (SP/SA) is a simple topology
consisting of a single attacker node sending packets along a single
identical path to a single victim node. The length of the path varies
with each work ranging from 3 hops to 32 hops \cite{savage,wong,ankunda,tseng}.
This setup is used to simulate the performance of PPM schemes during
a flooding style DoS attack.

The \textit{Single Path, Multiple Attacker} (SP/MA), and \textit{Multiple
Paths, Multiple Attacker} (MP/MA) topologies consist of multiple sources
of attack traffic to simulate a DDoS attack. The SP/MA simulates a
unique topology in which all the attackers are located at different
distances from the victim but all along a single identical path \cite{ankunda}.
The MP/MA simulates a more general topology where each attacker has
a unique path linking it to the victim node. In some cases, the paths
are completely independent \cite{wong}, while in other cases, the
paths merge closer to the victim \cite{ma,wong,goodrich,ankunda}.

One unique MP/MA topology is a tree, e.g. a binary tree. In this case,
the attack graph is modeled as a tree with some or all the leaves
at a certain depth representing the attack nodes, and the root of
the tree representing the victim node \cite{ma,wong,goodrich}. This
setup ensures that different attack paths merge the closer they are
to the victim. As with SP/SA and SP/MA, there is only one path in
the attack graph from an attack node to the victim node.

Schemes such as AMS, FIT, and TPM have been evaluated using actual
data sets from the Internet \cite{song,yaar1,paruchuri} including
traceroute data sets from Lucent Bell labs \cite{song} and CAIDA's
skitter map \cite{yaar1,paruchuri}. These data sets are used to produce
topologies that are typically larger than the simple topologies mentioned
thus far and provide better abstractions of the Internet structure.
However, they do not provide an adequate description of the Internet
since they are typically restricted to restricted snapshot of the
Internet network \cite{sung}. Traceroute datasets have also been
reported to be inaccurate when the network has asymmetric paths \cite{yaar1}. 

One common feature with these underlying topologies is their tree-like
structure. A tree-structured topology $G_{tree}$ exhibits a single
path from any given attacker to the victim. The choice of tree-structured
topologies is based on the assumption that all attack traffic from
one attacker will take the same path to the victim. This assumption
is in turn based on the observation that Internet paths are largely
invariant particularly over short periods of time \cite{paxson}.
These assumptions have allowed researchers to simplify the simulation
process by ignoring the routing capabilities of the network and enforcing
a predefined set of paths for attack traffic as evidenced in a tree-structured
topology. However, because tree-structured topologies do not exhibit
the alternative routes ubiquitous in the Internet, they can not be
used as an appropriate abstraction of the Internet topology. 

In summary, network models used to simulate the Internet topology
fall into three categories based on the properties that they emphasize,
namely degree-based models, structural models and spatial models (cf.
Fig. \ref{Internet}). The emphasis of degree-based models is the
degree distribution of the nodes in an attempt to recreate the power
law observations in the Internet \cite{faloutsos,waxman2}. The structural
models arrange the nodes to mimic the hierarchical structure of the
Internet, with Internet traffic being transmitted through routers
located within autonomous systems \cite{calvert,brite2}. The spatial
models place emphasis on the location of the nodes with any two nodes
being connected only if they are within a transmission range of each
other \cite{clark}.

\section{Topology Influence Analysis: A Demonstration\label{analysis}}

In the last part, we argued that the comparison of performance results
is difficult because the underlying topologies used are fundamentally
different. Here, an analytical basis is given that shows how subtle
differences in the network topologies contribute to differences between
the convergence times of the PPM-based schemes. 

The convergence time for PPM-based schemes has been originally modeled
as the \textit{coupon-collector's problem} \cite{feller,savage}.
A coupon collector seeks to collect $d$ equally likely distinct coupons
by drawing them from an urn with replacement. While it takes a short
time to get the first few unique coupons, it takes considerably longer
to get the last few coupons that complete the entire collection. The
expected number of turns needed to draw all $d$ distinct coupons
grows as $\Theta(d\cdot\ln(d))$ \cite{feller}.

\begin{figure}[!t]
\centering \subfigure{\label{attackgraphwithmot}\includegraphics[width=40mm]{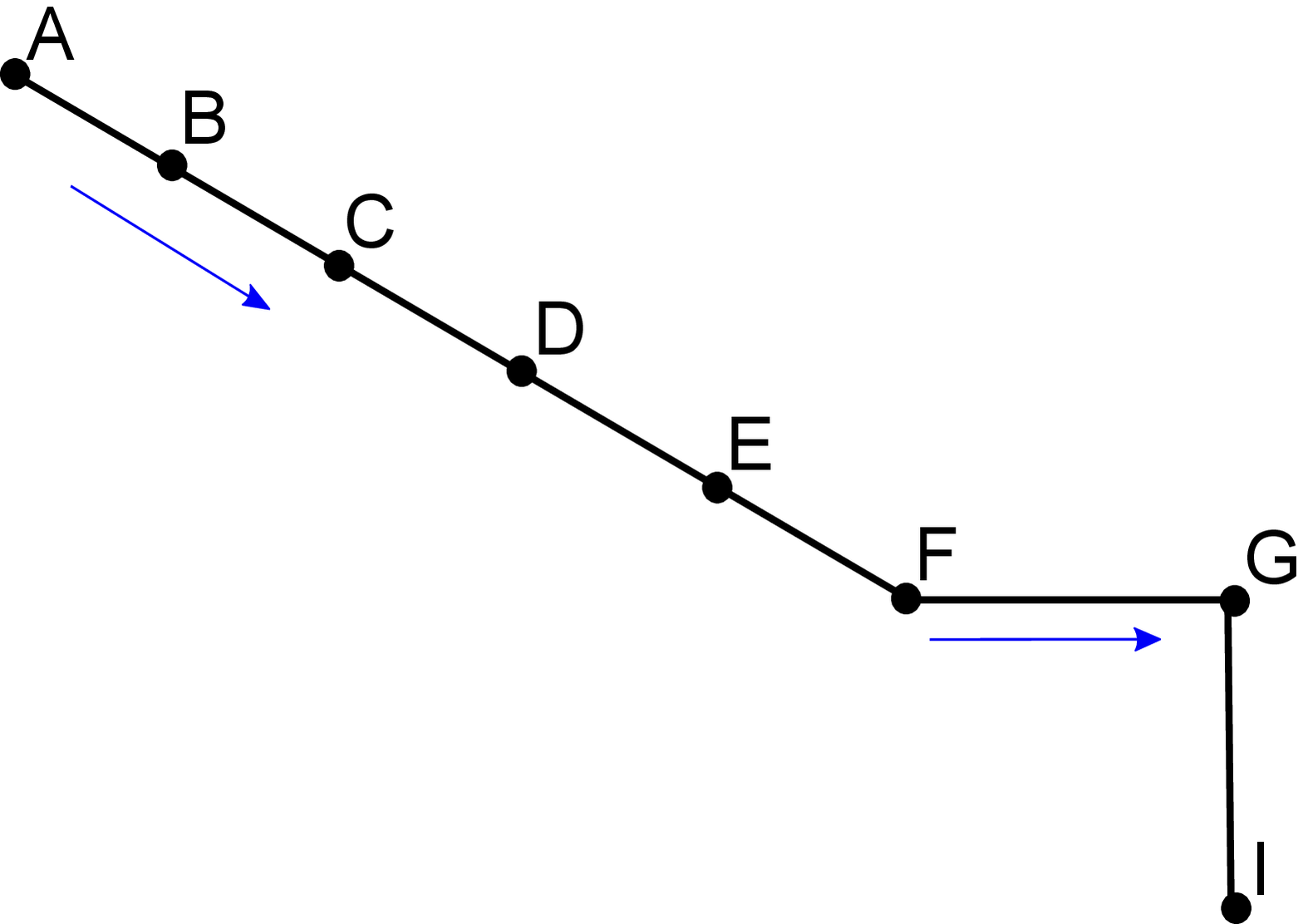}}\qquad{}\subfigure{\label{attackmot4}\includegraphics[width=40mm]{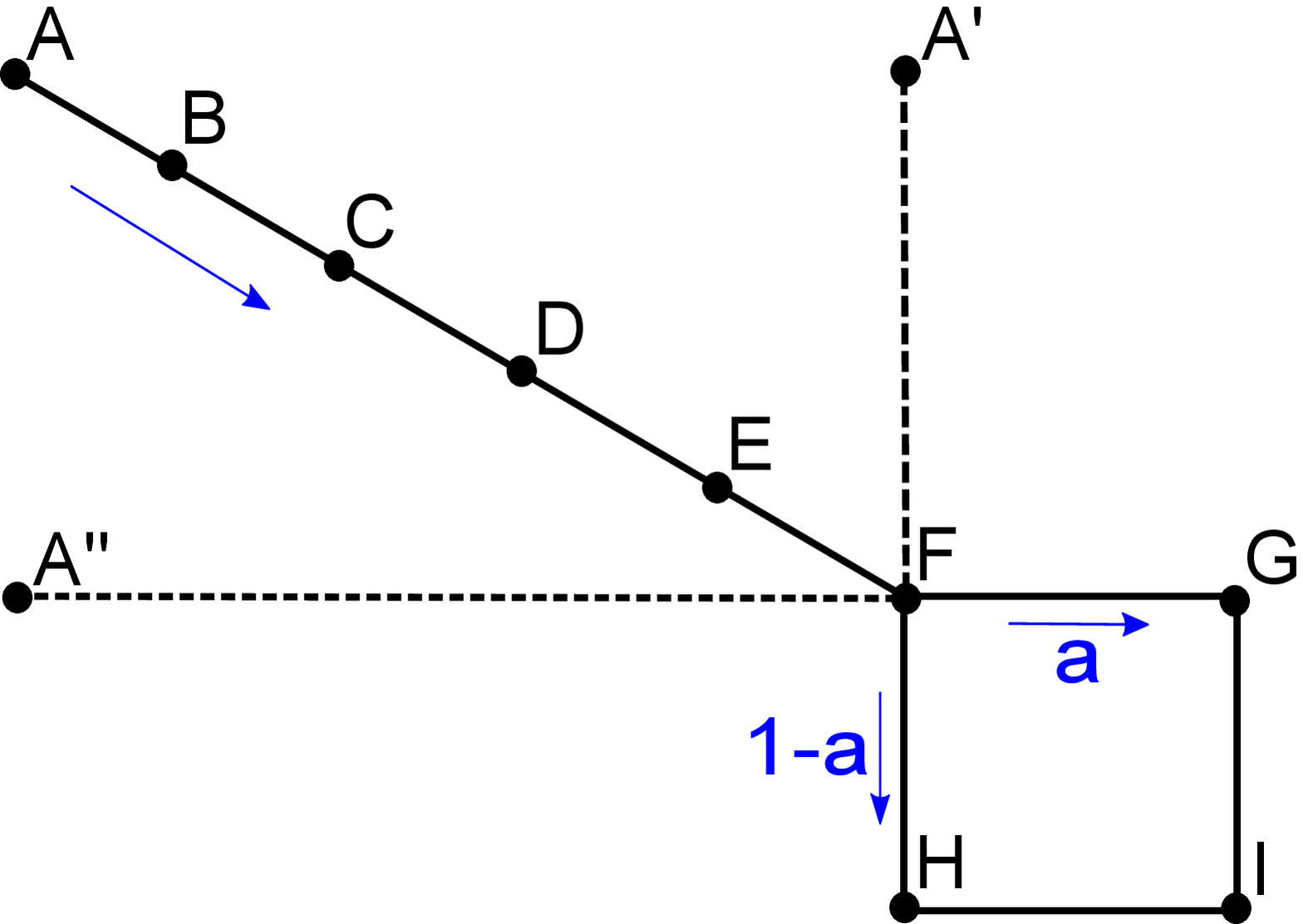}}
\caption{Sample attack paths linking attacker A to victim I. The attack path
in Fig. b exhibits Subgraph 4 in which the traffic can either take
path FGI with probability $a$, or path FHI with probability $1-a$.}
\label{attackmot} 
\end{figure}

When the coupon collector problem is applied to packet marking, the
marked packets are considered to be the coupons. For example, Fig.
\ref{attackgraphwithmot} shows a single path linking attacker A to
victim I and the target of the ``coupon collector'' would be to
collect markings for all 7 edges. However, the expected time expression
above must be modified to apply them to the packet marking problem
because one may or may not ``draw'' a marked packet in the packet
marking problem and the marked packets have unequal chances of being
received. Savage et al. \cite{savage} deal with the unequal edge
probabilities by utilizing the probability of the \textit{least likely
edge} to provide an upper bound on the expected convergence time.

Formally, given a single path of $l$ hops implementing the PPM scheme
with router marking probability $p$, the \textit{least likely edge}
is typically the edge located closest to the attacker which has a
probability $p(1-p)^{l-1}$ of being received by the victim. Given
$d$ unique markings, the probability of receiving any marking at
the victim is therefore at least $dp(1-p)^{l-1}$ which is the product
of the number of unique markings and the probability of the \textit{least
likely edge}. The expected number of packets $E[x]$ required to complete
the marking ``collection'' in order to build the attack graph is
derived by dividing the original coupon collector expectation by $dp(1-p)^{l-1}$
which yields \cite{savage}. 
\begin{equation}
E_{0,PPM}[x]<\frac{\ln(d)}{p(1-p)^{l-1}}\label{eq1}
\end{equation}

Therefore, the final expression for the upper bound of the expected
convergence time is obtained by dividing the natural logarithm of
the number of distinct edges $d$, by the probability $p(1-p)^{l-1}$
of the \textit{least likely edge} in the attack path. For the SP/SA
topology, the number of hops is equal to the number of unique markings
($l=d$).

In Fig. \ref{attackgraphwithmot}, the least likely edge is AB with
a probability of $p(1-p)^{l-1}$. However, the probability of receiving
edge FG in Fig. \ref{attackmot4} is given by $ap(1-p)$ which is
considerably less than the probability of AB for short path lengths.
In this case, the convergence time is given by Equation \ref{eq4}.
\begin{equation}
E_{1,PPM}[x]<\frac{\ln(d)}{ap(1-p)}\label{eq4}
\end{equation}
Comparing Equation \ref{eq1} and Equation \ref{eq4} reveals that
the convergence time is increased by a factor of $\frac{(1-p)^{l-2}}{a}$.
This means that even in the best case when both alternative paths
are equally likely ($a=0.5$), the convergence time of a 3-hop attack
graph is multiplied by a factor of $1.92$ while a 15 hop attack graph
is multiplied by a factor of $1.18$. If one of the two paths only
carries a tenth of the traffic ($a=0.1$), the convergence times of
the 3-hop and 15-hop attack graphs are multiplied by a factor of $9.6$
and $5.88$ respectively. Additionally, the alternative paths factor
affects short attack paths more than long attack paths.

The cases for TMS and PBS are similar. It can be shown that for TMS
and PBS the convergence time is increased by a factor of $\frac{1}{a}$
regardless of the path length. 

This shows that alternative paths reduce the probability of the least
likely edges in an attack path for all the considered schemes, and
consequently increases their convergence times. 

\section{Conclusion\label{sec:Conclusion}}

In this paper, ten PPM-based IP traceback schemes are compared and
analyzed. Although, it is a non-exhaustive study, the number of schemes
is sufficiently representative to show existing discrepancies in performance
and performance evaluation between the schemes.

As a result of this analysis, it becomes comprehensive why a direct
comparison of PPM-based schemes turns out to be a prohibitively complicated
undertaken. The reasons for this: Firstly, most schemes utilize different
metrics to measure their performance. Secondly, many schemes are evaluated
on different kinds of underlying network topologies, the majority
of which do not provide an adequate abstraction of the topology of
the Internet or focus on different characteristics of it. 

The underlying topologies are typically tree-structured with a single
path from an attacker to the victim. In contrast, the topology of
the Internet exhibits alternative routes that make it both resilient
and scalable. Both the disparity in metrics and underlying topologies,
as well as the inadequacy of the topologies, raise questions about
the performances of these schemes in the Internet. For example, which
scheme would perform best when deployed on an appropriate network\textemdash a
network similar to the Internet? Does the performance of a PPM-based
scheme vary with the network on which it is implemented? If so, is
it possible to provide a common ground for comparing scheme performance
when the schemes are implemented on different networks? These questions
show that there is a need to evaluate and compare the schemes on common
and appropriate networks. The results of this evaluation can then
be used to determine which schemes are the most promising candidates
for Internet deployment.

The work presented herein has implications that reach outside the
field of IP traceback. For example, do other network protocols also
display a dependence on the network topology exhibited by the network
on which they are implemented? If so, can these dependencies be exploited
to yield better protocol performance in specific types of networks?
These questions should encourage multiple topology evaluations of
network protocols.

\balance

\bibliographystyle{ieeetr}
\bibliography{IPtraceback23}

\end{document}